\documentstyle[twoside,fleqn,npb,epsfig]{article}

\title{$\sigma(e^+e^-\to\mbox{hadrons})$ and the Heavy Quark Masses}

\author{M. Steinhauser\address{II. Institut f\"ur Theoretische Physik,
    Universit\"at Hamburg, \\
    Luruper Chaussee, D-22761 Hamburg, Germany}%
  \thanks{Talk given at RADCOR/Loops and Legs, September 2002.}}

\begin{document}

\begin{abstract}
The precise data for the total cross
section $\sigma(e^+e^-\to\mbox{hadrons})$
from the charm threshold region,
when combined with the evaluation of moments with three loop
accuracy, lead to a direct determination of the short distance
$\overline{\rm MS}$ charm quark mass $m_c(m_c)=1.304(27)$.
Applying the
same approach to the bottom quark we obtain $m_b(m_b)=4.191(51)$~GeV. 
A complementary method for the determination of $m_b$ is based of the analysis
of the $\Upsilon(1S)$ system which is confronted with a
next-to-next-to-next-to-leading order calculation of the corresponding
energy level. This leads to $m_b(m_b)=4.346(70)$~GeV.
\end{abstract}

\maketitle

%%%%%%%%%%%%%%%%%%%%%%%%%%%%%%%%%%%%%%%%%%%%%%%%%%%%%%%%%%%%

\section{Introduction}

During the past years new and more precise data for 
$\sigma(e^+e^-\to\mbox{hadrons})$ 
have become available in the low energy region between 2 and 10~GeV. 
At the same time increasingly precise calculations have been
performed in the framework of perturbative QCD (pQCD), 
both for the cross section as a
function of the center-of-mass energy $\sqrt{s}$,
including quark mass effects, and for its moments which
allow for a precise determination of the quark mass.
A fresh look at
the evaluation of the charm quark mass with the help of sum rules is
thus an obvious task.
We will concentrate on low moments
as suggested by the ITEP group long ago~\cite{NSVZ}.
This is a natural route to determine directly a short distance
mass, say $m_c(m_c)$, in the 
$\overline{\rm MS}$ scheme~\cite{NSVZ,ShiVaiZak79}.

The same method can also be applied to the bottom system leading directly to
$m_b(m_b)$.

A complementary method for the determination of $m_b$ is based of the analysis
of the $\Upsilon(1S)$ system.
Recently a major improvement on the theoretical side has been achieved by the
evaluation of the next-to-next-to-next-to-leading order 
(N$^3$LO) corrections to the energy level. The disadvantage of
this method is due to the large non-perturbative effects.

%%%%%%%%%%%%%%%%%%%%%%%%%%%%%%%%%%%%%%%%%%%%%%%%%%%%%%%%%%%%

\section{\label{sec::mcmb}Charm 
  and bottom quark mass from low-order moment sum rules}

\begin{figure*}[htb]
  \begin{center}
    \begin{tabular}{c}
      \leavevmode
      \epsfxsize=14cm
     \epsffile[40 250 550 580]{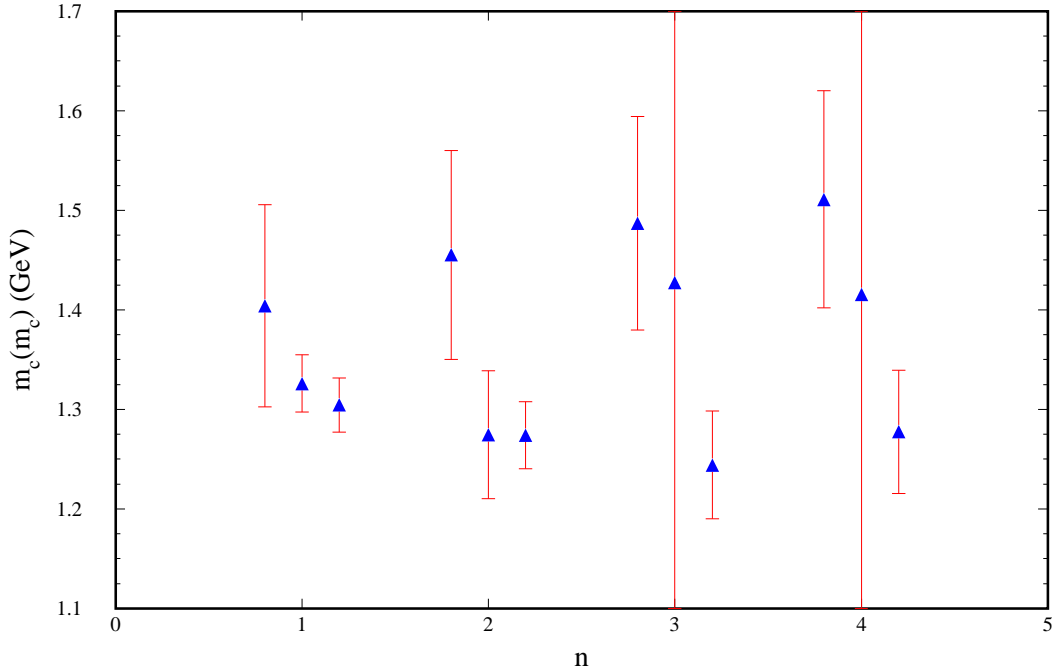}
    \end{tabular}
  \end{center}
  \vspace{-3.em}
  \caption{\label{fig:mom}$m_c(m_c)$ for $n=1,2,3$ and $4$.
  For each value of $n$ the results from left to right correspond to
  the inclusion of terms of order $\alpha_s^0$, $\alpha_s^1$ and
  $\alpha_s^2$ in the coefficients $\bar{C}_n$.
  Note, that for $n=3$ and $n=4$ the errors can not be determined
  with the help of Eq.~(\ref{eq:mc1})
  in those cases where only the two-loop corrections of order $\alpha_s$ are
  included into the coefficients $\bar{C}_n$ as the equation cannot be
  solved for $m_c$.
  }
\end{figure*}

A very elegant method for the determination of the charm quark
mass is based on the direct comparison of theoretical and experimental
moments of the charm quark contribution to the photon polarization
function.
In the limit of small momentum the latter can be cast 
into the form~\cite{CheKueSte96}
\begin{eqnarray}
  \Pi_c(q^2) &=& Q_c^2 \frac{3}{16\pi^2} \sum_{n\ge0}
                       \bar{C}_n z^n
  \,,
  \label{eq:pimom}
\end{eqnarray}
with 
$z=q^2/(4m_c^2)$ where $m_c=m_c(\mu)$ is the $\overline{\rm MS}$ 
charm quark mass at the scale $\mu$.
The perturbative series for the coefficients $\bar{C}_n$ 
up to $n=8$ is known
analytically~\cite{CheKueSte96} up to order $\alpha_s^2$.
We define the moments through
\begin{eqnarray}
  {\cal M}_n^{\rm exp} &=& \int \frac{{\rm d}s}{s^{n+1}} R_c(s)
  \nonumber\\
  &=&
  {\cal M}_n^{\rm th} \,\,=\,\, 
  \frac{9}{4}Q_c^2
  \left(\frac{1}{4 m_c^2}\right)^n \bar{C}_n
  \,,
  \label{eq:Mth}
\end{eqnarray}
which leads to the following formula for the charm quark mass
\begin{eqnarray}
  m_c(\mu) &=& \frac{1}{2} 
  \left(\frac{\bar{C}_n(\ln m_c)}{{\cal M}_n^{\rm exp}}\right)^{1/(2n)} 
  \,.
  \label{eq:mc1}
\end{eqnarray}
The analysis of the experimental moments with $n=1,\ldots,4$ leads to
the results displayed in Fig.~\ref{fig:mom}~\cite{KueSte01}.

The moment with $n=1$ is evidently least sensitive to non-perturbative
contributions from condensates, to the Coulombic higher order effects, the
variation of $\mu$ and the parametric $\alpha_s$ dependence.
Hence 
\begin{eqnarray}
  m_c(m_c) &=& 1.304(27)~\mbox{GeV}
  \,.
  \label{eq:mcfinal}
\end{eqnarray}
is adopted as the final result~\cite{KueSte01}.
In principle the same analysis can be performed using the pole mass sheme for
the quarks. However, the final results are quite unstable in contrast to the 
$\overline{\rm MS}$ scheme.

The same approach is also applicable to the determination of
$m_b$.
Again a significant improvement of the stability of the prediction 
after inclusion of the NNLO terms is observed.
As final result one finds
\begin{eqnarray}
  m_b(m_b) &=& 4.191(51)~\mbox{GeV}
  \,,
  \label{eq:mbmb}
\end{eqnarray}

%%%%%%%%%%%%%%%%%%%%%%%%%%%%%%%%%%%%%%%%%%%%%%%%%%%%%%%%%%%%

\section{The $\Upsilon(1S)$ system and the bottom quark mass}

In contrast to the previous section one has to deal with a
non-relativistic system of a bound state of a heavy quark-antiquark
pair which is governed by
a complicated multiscale dynamics.
In the nonrelativistic regime, where the heavy-quark velocity $v$ is of the
order of the strong-coupling constant $\alpha_s$, the Coulomb effects are
crucial and have to be taken into account to all orders in $\alpha_s$.
This makes the use of the effective theory mandatory.
This approach allows us to separate the scales and to
implement the expansion in $v$ at the level of the Lagrangian.

The dynamics of a nonrelativistic quark-antiquark pair is
characterized by four different regions,
the hard region, 
the soft region,
the potential region,
and the ultrasoft region.
Nonrelativistic QCD (NRQCD) \cite{CasLep} is obtained by integrating out
the hard modes.
Subsequently integrating out the soft modes and the potential gluons results
in the effective theory of potential NRQCD (pNRQCD) \cite{PinSot1}, which
contains potential heavy quarks and ultrasoft gluons, ghosts, and light quarks
as active particles.
The effect of the modes that have been integrated out is two-fold:
higher-dimensional operators appear in the effective Hamiltonian,
corresponding to an expansion in $v$, and the Wilson coefficients of the
operators in the effective Hamiltonian acquire  corrections, which are series
in $\alpha_s$.
In~\cite{KPSS1,KPSS2} the ingredients for the N$^3$LO Hamiltonian have been
completed using an efficient combination of the effective theory formalism and
the threshold expansion~\cite{BenSmi}.

Once the Hamiltonian is available the only task is to solve the corresponding
Schr\"odinger equation up to third order using the usual time-independent
perturbation theory. 
A new feature which appears for the first time at N$^3$LO are the retardation
effects which 
arise from the chromoelectric dipole interaction of the heavy quarkonium
with a virtual ultrasoft gluon.

The third order correction to the lowest energy level 
was calculated in~\cite{KPSS2,PS1} and will be used 
in this work in order to determine the bottom quark mass.
The application to the top
quark system can be found in~\cite{Peninproc}.
Recently the N$^3$LO Hamiltonian has been used to compute 
the corrections of order $\alpha_s^3\ln\alpha_s$ to the wave
function~\cite{KPSS3}.

The perturbative expansion of the energy level with quantum number $n$ looks
as follows
\begin{eqnarray}
  E_n^{\rm p.t.}=E^C_n+\delta E^{(1)}_n+\delta E^{(2)}_n+\delta
  E^{(3)}_n
  +\ldots\,,
\end{eqnarray}
where 
the ${\cal O}\left(\alpha_s^3\right)$ correction, $E^{(3)}_n$, 
arises from the following sources:
\begin{itemize}
\item[(i)] matrix elements of the N$^3$LO operators of the effective
Hamiltonian between Cou\-lomb wave functions; 
\item[(ii)] higher iterations of the NLO and NNLO operators of the effective
Hamiltonian in time-independent perturbation theory; 
\item[(iii)] matrix elements of the N$^3$LO instantaneous operators generated
by the emission and absorption of ultrasoft gluons; and
\item[(iv)] the retarded ultrasoft contribution.
\end{itemize}

\begin{figure}[ht]
  \begin{center}
    \begin{tabular}{c}
      \leavevmode
      \epsfxsize=7.2cm
      \epsffile[40 230 550 580]{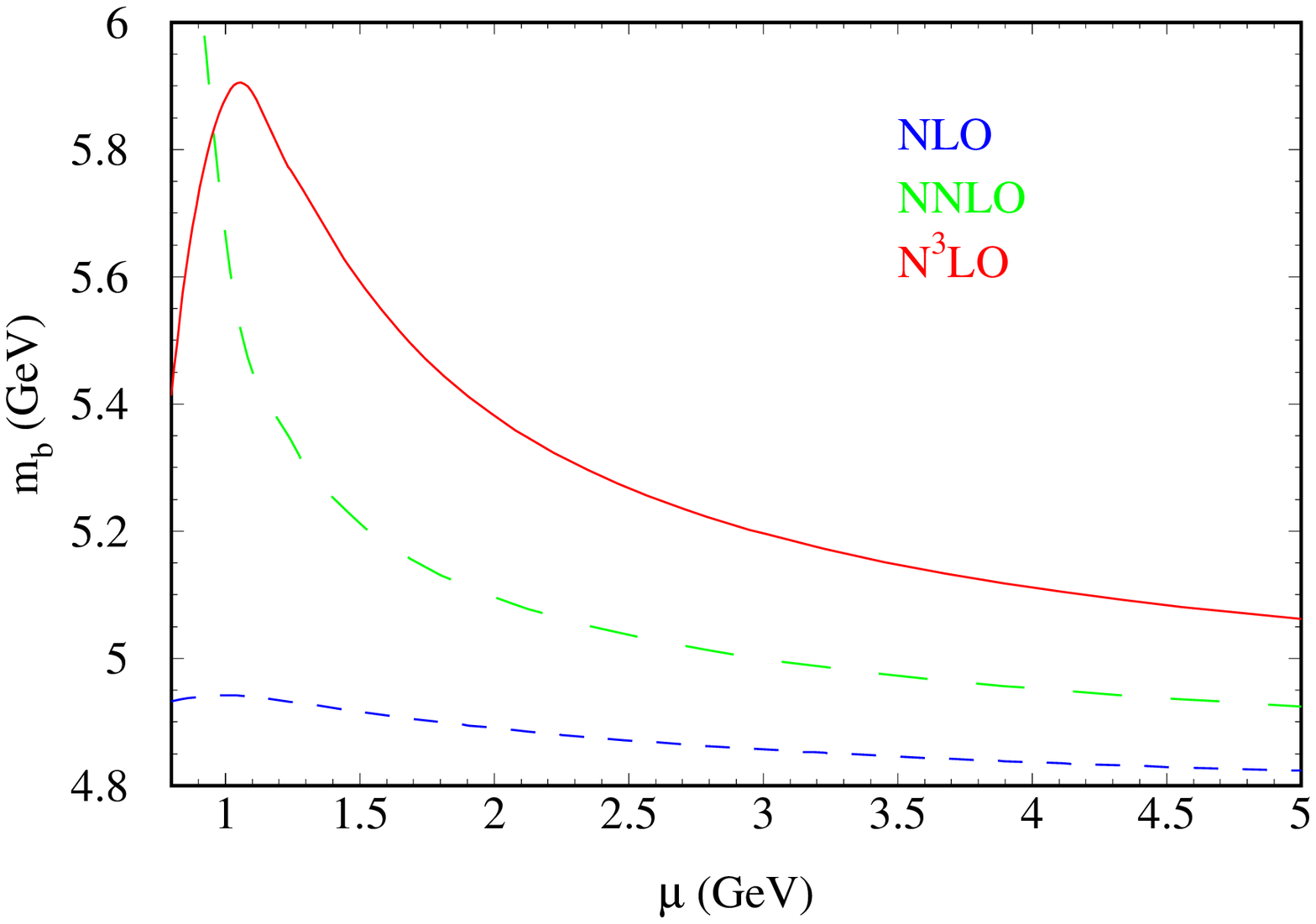}
      \\
      (a)
      \\
      \leavevmode
      \epsfxsize=7.2cm
      \epsffile[40 230 550 580]{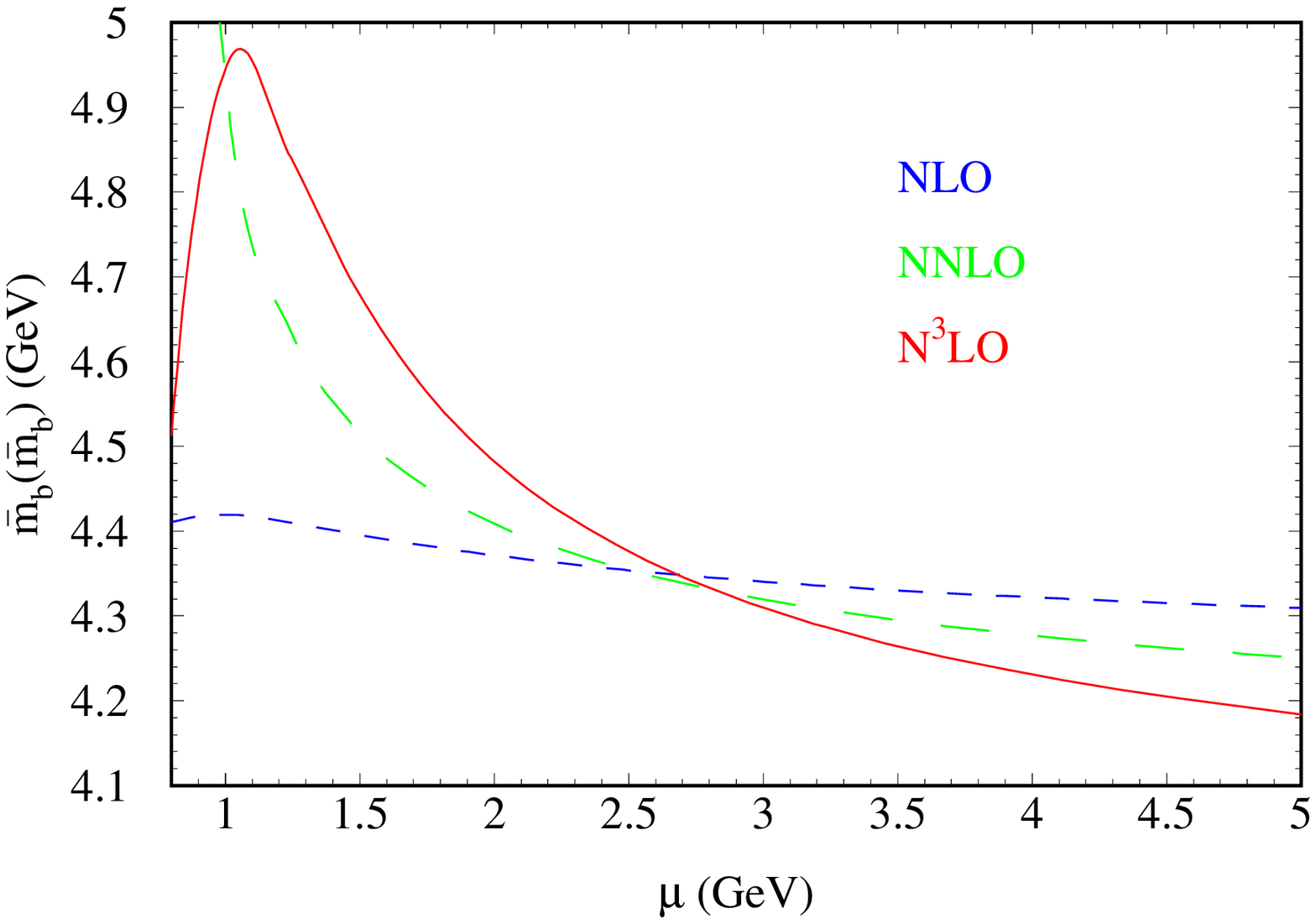}
      \\
      (b)
    \end{tabular}
  \end{center}
  \vspace{-2em}
  \caption{\label{fig:botpol}
    (a) Bottom quark pole mass, $m_b$, as a function
    of the renormalization scale $\mu$, used for the numerical
    evaluation of $E_1$ where the short-dashed, long-dashed and solid line
    corresponds to the NLO, NNLO and N$^3$LO approximations.
    (b) $\overline{\rm MS}$ bottom quark mass
    $\overline{m}_b(\overline{m}_b)$, 
    as a function
    of the renormalization scale $\mu$, which is used 
    for the extraction of the pole mass $m_b$ (cf. (a)). For the
    conversion from the pole to the $\overline{\rm MS}$ mass the
    method described around Eq.~(\ref{mustar}) is used.
          }
\end{figure}

The analytical result for $E^{(3)}_n$
can be found in~\cite{PS1}. In numerical form
it reads (adopting the choice $\mu_s=C_F\alpha_s(\mu_s) m_q$)
for the bottom and top system
\begin{eqnarray}
  \delta E^{(3)}_1&=& \alpha_s^3(\mu_s)E_1^C\left[
  \left(
  \begin{array}{c} 
    70.590|_{\mbox{\tiny bottom}}\\
    56.732|_{\mbox{\tiny top}} 
  \end{array} 
  \right)
  \right.
  \nonumber\\&&
  \left.
  + 15.297 \ln(\alpha_s(\mu_s)) + 0.001\,a_3 
  \right.
  \nonumber\\&&
  \left.
  + \left(
  \begin{array}{c} 
    34.229|_{\mbox{\tiny bottom}}\\
    26.654|_{\mbox{\tiny top}} 
  \end{array} 
  \right)\bigg|_{\beta_0^3}
  \right]\,,
  \label{eq:dele3num}
\end{eqnarray}
where we have separated the contributions arising from $a_3$ and
$\beta_0^3$.
The only unknown ingredient in the result for $\delta E^{(3)}_1$
is the three-loop $\overline{\rm MS}$ coefficient $a_3$ 
of the corrections to the static potential.
Up to now there are only estimates based on Pad\'e approximation~\cite{ChiEli}
which is used in our analysis.
However, the final result only changes marginally
even for a rather large deviations of $a_3$ from its  Pad\'e estimate.

The starting point for the determination of the bottom quark mass
is its relation to the mass of the $\Upsilon(1S)$  
resonance
\begin{eqnarray}
  M_{\Upsilon(1S)} = 2m_b + E_1^{\rm p.t.}+ \delta^{\rm n.p.}E_1\,,
  \label{eq:M1SE}
\end{eqnarray}
with $M_{\Upsilon(1S)}=9.46030(26)$~GeV.
Here $\delta^{\rm n.p.}E_1$ is the nonperturbative correction
to the ground state energy. 
The dominant contribution is associated with the
gluon condensate and gives~\cite{PS1}
\begin{equation}
  \delta^{\rm n.p.}E_1=60\pm 30~{\rm MeV}\,.
  \label{nptb}
\end{equation}

Combining  Eq.~(\ref{eq:M1SE})  with the result for the perturbative
perturbative ground state energy up to ${\cal O}(m_q\alpha_s^5)$
one obtains the bottom quark mass as a function the renormalization
scale of the strong coupling constant normalization, $\mu$,
which is plotted in Fig.~\ref{fig:botpol}(a).
For the numerical evaluation we extract 
$\alpha_s^{(4)}(m_b)$ with $m_b=4.83$~GeV from its value at $M_Z$ using
four-loop $\beta$-function accompanied with three-loop
matching\footnote{We use the package {\tt
    RunDec}~\cite{Chetyrkin:2000yt} to perform the running and
  matching of $\alpha_s$}.
$\alpha_s^{(4)}(m_b)$ is used as starting point in order to evaluate
$\alpha_s^{(4)}(\mu)$ at N$^k$LO
with the help of the $k$-loop $\beta$ function.
From Fig.~\ref{fig:botpol}(a) we see that the dependence on the
renormalization scale
becomes very strong below $\mu\sim 2$~GeV which 
indicates that the perturbative corrections are not under control. 
However, even above this scale
the perturbative series for the pole mass shows no sign of convergence.
This means that one can assign a numerical value to the  
pole mass only in a specified order of perturbation theory. 

On the contrary, it is widely believed that the 
$\overline{\rm MS}$ mass ${\overline m}_b(\mu)$
at the scale $\mu={\overline m}_b(\mu)$ is a short-distance object
which has much better perturbative properties. 
Thus, it seems to be reasonable to convert our result for the pole 
mass into ${\overline m}_b({\overline m}_b)$. The relation between $m_b$ 
and ${\overline m}_b({\overline m}_b)$ is known up to 
three-loop approximation
\cite{CheSte99,MelRit99} and shows
sizable perturbative corrections.
For this reason we suggest the following procedure to take
into account these corrections in a most accurate way.  
The idea is that from the one-parametric family
of ${\overline m}_b(\mu)$
we can choose a representative corresponding to some scale $\mu^\star$ 
in such a way that
\begin{equation}
  {\overline m}_b(\mu^\star)=m_b\,.
  \label{mustar}
\end{equation}
For a given fixed-order value of the pole mass
Eq.~(\ref{mustar}) can be solved for $\mu^\star$.
In particular, for a pole mass to N$^k$LO 
we use the $k$-loop relation between the $\overline{\rm MS}$ 
and pole mass in Eq.~(\ref{mustar}).
Afterwards ${\overline m}_b({\overline m}_b)$ can be computed from
$\overline{m}_b(\mu^\star)$ solving the renormalization group (RG) equation.
The advantage of this approach is obvious: we use the finite order relation 
between  $\overline{\rm MS}$ and pole mass at the scale
where they are perturbatively close while the
large difference between ${\overline m}_b({\overline m}_b)$
and $m_b$ is completely covered  by the RG evolution 
which can be computed with very high accuracy as the corresponding  
anomalous dimension is known to four-loop approximation  
\cite{Chetyrkin:1997dh,Vermaseren:1997fq}. The only restriction
on the method could be connected to the value of $\mu^\star$. 
It should be large enough to allow for a reliable use of the RG
equation which in practice indeed is the case~\cite{PS1}.

In Fig.~\ref{fig:botpol}(b) 
our result for  ${\overline m}_b({\overline m}_b)$  
is plotted at NLO, NNLO and N$^3$LO
as a function of the normalization scale $\mu$ 
which is used to obtain the pole mass $m_b$
(cf. Fig.~\ref{fig:botpol}(a)).
It is remarkable that close to $\mu= 2.7$~GeV, which is consistent
with the physically motivated soft scale $\mu_s\approx 2$~GeV, 
both the second and the third order corrections vanish.
This fact is a rather strong indication of the 
convergence of the series for ${\overline m}_b({\overline m}_b)$.

The uncertainties in the obtained value of 
${\overline m}_b({\overline m}_b)$ have been investigated in detail
in~\cite{PS1}.
Finally, the prediction for the $\overline{\rm MS}$ 
bottom quark mass reads
\begin{equation}
  {\overline m}_b({\overline m}_b)=4.346\pm 0.070~{\rm GeV}
  \,.
  \label{finalmb1}
\end{equation}

%%%%%%%%%%%%%%%%%%%%%%%%%%%%%%%%%%%%%%%%%%%%%%%%%%%%%%%%%%%%

\section{Conclusions}

For the comparison of the results discussed in this contribution
with the literature we refer to~\cite{KueSte01,PS1}.
It is, however, interesting to compare the value for the charm quark
mass with a very recent result obtained in a lattice
calculation~\cite{Rolf:2002gu}. Their final result in quenched
approximation, $m_c(m_c)=1.301(34)$~GeV,
is impressively close to ours (cf. Eq.~(\ref{eq:mcfinal}))
with comparable errors. In order to estimate the uncertainty 
induced by the
quenched approximation we performed a ``perturbative quenching'' by
setting the number of active
flavours to zero in the calculation of~\cite{KueSte01}. 
This leads to a small uncertainty of $\approx30$~MeV,
which agrees with the estimate of~\cite{Rolf:2002gu}.

Although formally slightly beyond 1$\sigma$ the $\overline{\rm MS}$
quark mass obtained from the low-moment sum rule approach is in very
good agreement with the one from the $\Upsilon(1S)$ system. One needs
to have in mind that both the experimental data and the theoretical
calculations are completely different.
Whereas no further improvement in the $\Upsilon(1S)$ method can be
expected there is significant improvement possible in the 
approach discussed in Section~\ref{sec::mcmb}. In particular,
after reducing the ex\-perimental
error of $R(s)$
in the charm and bottom threshold region and the one of the
leptonic widths of the narrow resonances to roughly 2\%
a reduction of the uncertainty in the charm and bottom quark mass to 
15~MeV and 30~MeV, respectively, can be expected. More
details can be found in Ref.~\cite{Kuhn:2002zr}.

%%%%%%%%%%%%%%%%%%%%%%%%%%%%%%%%%%%%%%%%%%%%%%%%%%%%%%%%%%%%

\vspace{-.2em}

\section*{Acknowledgment}

I would like to thank the organizers of RADCOR/Loops and Legs 
2002 conference for the kind
inviation and B.A. Kniehl, J.H. K\"uhn, A.A. Penin and V.A. Smirnov for a
fruitful collaboration.

%%%%%%%%%%%%%%%%%%%%%%%%%%%%%%%%%%%%%%%%%%%%%%%%%%%%%%%%%%%%


\begin{thebibliography}{99}

\def\npb#1#2#3{  {Nucl.\ Phys.\ B }{\bf #1} (#2) #3}
\def\plb#1#2#3{  {Phys.\ Lett.\ B }{\bf #1} (#2) #3}
\def\prl#1#2#3{  {Phys.\ Rev.\ Lett.\ }{\bf #1} (#2) #3}
\def\zpc#1#2#3{  {Z.\ Phys.\ C }{\bf #1} (#2) #3}

\bibitem{NSVZ} 
V.~A.~Novikov {\em et al.}, Phys.\ Rep.\ C {\bf 41} (1978) 1.

\bibitem{ShiVaiZak79}
M.~A.~Shifman, A.~I.~Vainshtein, and V.~I.~Zakharov,
\npb{147}{1979}{385}; \npb{147}{1979}{448}.

\bibitem{CheKueSte96}
  K.~G. Chetyrkin, J.~H. K\"uhn, and M. Steinhauser, \plb{371}{1996}{93};
  \npb{482}{1996}{213}; \npb{505}{1997}{40}.

\bibitem{KueSte01}
J.H. K\"uhn and M. Steinhauser,
Nucl.\ Phys.\ B {\bf 619} (2001) 588, (E) ibid. B {\bf 640} (2002) 415.

\bibitem{CasLep}
W.E. Caswell and G.P. Lepage,
Phys.\ Lett.\ B 167 (1986) 437;
G.T. Bodwin, E. Braaten, and G.P. Lepage,
Phys.\ Rev.\ D 51 (1995) 1125; (E) ibid. D 55 (1997) 5853.

\bibitem{PinSot1}
A. Pineda and J. Soto,
Nucl.\ Phys.\ B (Proc.\ Suppl.) 64 (1998) 428.

\bibitem{KPSS1}
B.A. Kniehl, A.A. Penin, V.A. Smirnov, and M. Steinhauser, 
Phys. Rev. D {\bf 65} (2002) 091503(R).

\bibitem{KPSS2}
B.A. Kniehl, A.A. Penin, V.A. Smirnov, and M. Steinhauser, 
Nucl.\ Phys.\ {\bf B635}, 357 (2002).

\bibitem{BenSmi}
M. Beneke and V.A. Smirnov,
Nucl.\ Phys.\ B 522 (1998) 321.

\bibitem{PS1}
A.A. Penin and M. Steinhauser, Phys. Lett. b {\bf 538} (2002) 335.

\bibitem{Peninproc}
A.A. Penin, these proceedings, hep-ph/0210201.

\bibitem{KPSS3}
B.A. Kniehl, A.A. Penin, V.A. Smirnov, and M. Steinhauser, 
Report No.\ DESY~02-134, hep-ph/0210161.

\bibitem{ChiEli}
F.A. Chishtie and V. Elias,
Phys.\ Lett.\ B 521 (2001) 434.

\bibitem{Chetyrkin:2000yt}
K.~G.~Chetyrkin, J.~H.~K\"uhn, and M.~Steinhauser,
%``RunDec: A Mathematica package for running and decoupling of the strong  coupling and quark masses,''
Comput.\ Phys.\ Commun.\  {\bf 133} (2000) 43.

\bibitem{CheSte99}
  K.~G. Chetyrkin and M. Steinhauser, \prl{83}{1999}{4001}; 
  \npb{573}{2000}{617}.
  
\bibitem{MelRit99}
  K.~Melnikov and T.~v.~Ritbergen, Phys.\ Lett.\ B {\bf 482} (2000) 99.

\bibitem{Chetyrkin:1997dh}
K.~G.~Chetyrkin,
%``Quark mass anomalous dimension to O(alpha(s)**4),''
Phys.\ Lett.\ B {\bf 404} (1997) 161.
%%%[arXiv:hep-ph/9703278].
%%CITATION = HEP-PH 9703278;%%

\bibitem{Vermaseren:1997fq}
J.~A.~Vermaseren, S.~A.~Larin, and T.~van Ritbergen,
%``The 4-loop quark mass anomalous dimension and the invariant quark  mass,''
Phys.\ Lett.\ B {\bf 405} (1997) 327.
%%%[arXiv:hep-ph/9703284].
%%CITATION = HEP-PH 9703284;%%

\bibitem{Rolf:2002gu}
J.~Rolf and S.~Sint  [ALPHA Collaboration],
%``A precise determination of the charm quark's mass in quenched QCD,''
Report No.: hep-ph/0209255.

\bibitem{Kuhn:2002zr}
J.H.~K\"uhn and M.~Steinhauser,
%``The impact of sigma(e+ e- $\to$ hadrons) measurements at intermediate  energies on the parameters of the standard model,''
JHEP {\bf 0210} (2002) 018.


\end{thebibliography}
\end{document}